\documentclass{article}

\usepackage{arxiv}

\usepackage[utf8]{inputenc} 
\usepackage[T1]{fontenc}    
\usepackage{hyperref}       
\usepackage{url}            
\usepackage{booktabs}       
\usepackage{amsfonts}       
\usepackage{nicefrac}       
\usepackage{microtype}      
\usepackage{lipsum}		
\usepackage{graphicx}
\usepackage{natbib}
\usepackage{doi}

\title{Phase Space Analysis of Cardiac Spectra}

\author{ \href{https://orcid.org/0000-0002-0082-8209}{\includegraphics[scale=0.06]{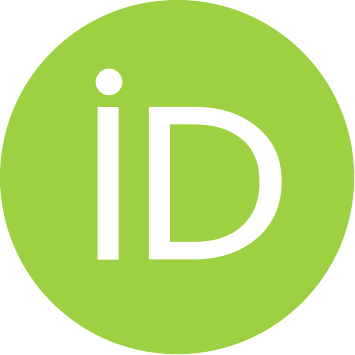}\hspace{1mm}Onder Pekcan} \\
	Department of Molecular Biology and Genetics\\
	Kadir Has University\\
	Istanbul, Turkey \\
	\texttt{pekcan@khas.edu.tr} \\
	\And
	\href{https://orcid.org/0000-0002-4453-3218}{\includegraphics[scale=0.06]{orcid.pdf}\hspace{1mm}Taner Arsan} \thanks{Corresponding Author}\\
	Department of Computer Engineering\\
	Kadir Has University\\
	Istanbul, Turkey \\
	\texttt{arsan@khas.edu.tr} \\
}

\renewcommand{\shorttitle}{\textit{arXiv} Template}

\hypersetup{
pdftitle={A template for the arxiv style},
pdfsubject={q-bio.NC, q-bio.QM},
pdfauthor={David S.~Hippocampus, Elias D.~Striatum},
pdfkeywords={First keyword, Second keyword, More},
}

\begin{document}
\maketitle

\begin{abstract}
	Cardiac diseases are one of the main reasons of mortality in modern, industrialized societies, and they cause high expenses in public health systems. Therefore, it is important to develop analytical methods to improve cardiac diagnostics. Electric activity of heart was first modeled by using a set of nonlinear differential equations. Latter, variations of cardiac spectra originated from deterministic dynamics are investigated. Analyzing the power spectra of a normal human heart presents His-Purkinje network, possessing a fractal like structure. Phase space trajectories are extracted from the time series graph of ECG. Lower values of fractal dimension, $D$ indicate dynamics that are more coherent. If $D$ has non-integer values greater than two when the system becomes chaotic or strange attractor. Recently, the development of a fast and robust method, which can be applied to multichannel physiologic signals, was reported. This manuscript investigates two different ECG systems produced from normal and abnormal human hearts to introduce an auxiliary phase space method in conjunction with ECG signals for diagnoses of heart diseases. Here, the data for each person includes two signals based on $V_4$ and modified lead III (MLIII) respectively. Fractal analysis method is employed on the trajectories constructed in phase space, from which the fractal dimension $D$ is obtained using the box counting method. It is observed that, MLIII signals have larger $D$ values than the first signals ($V_4$), predicting more randomness yet more information. The lowest value of $D$ (1.708) indicates the perfect oscillation of the normal heart and the highest value of $D$ ( 1.863) presents the randomness of the abnormal heart. Our significant finding is that the phase space picture presents the distribution of the peak heights from the ECG spectra, giving valuable information about heart activities in conjunction with ECG.
\end{abstract}


\section{Introduction}
\begin{figure}
	\centering
	\includegraphics[width=16.5cm]{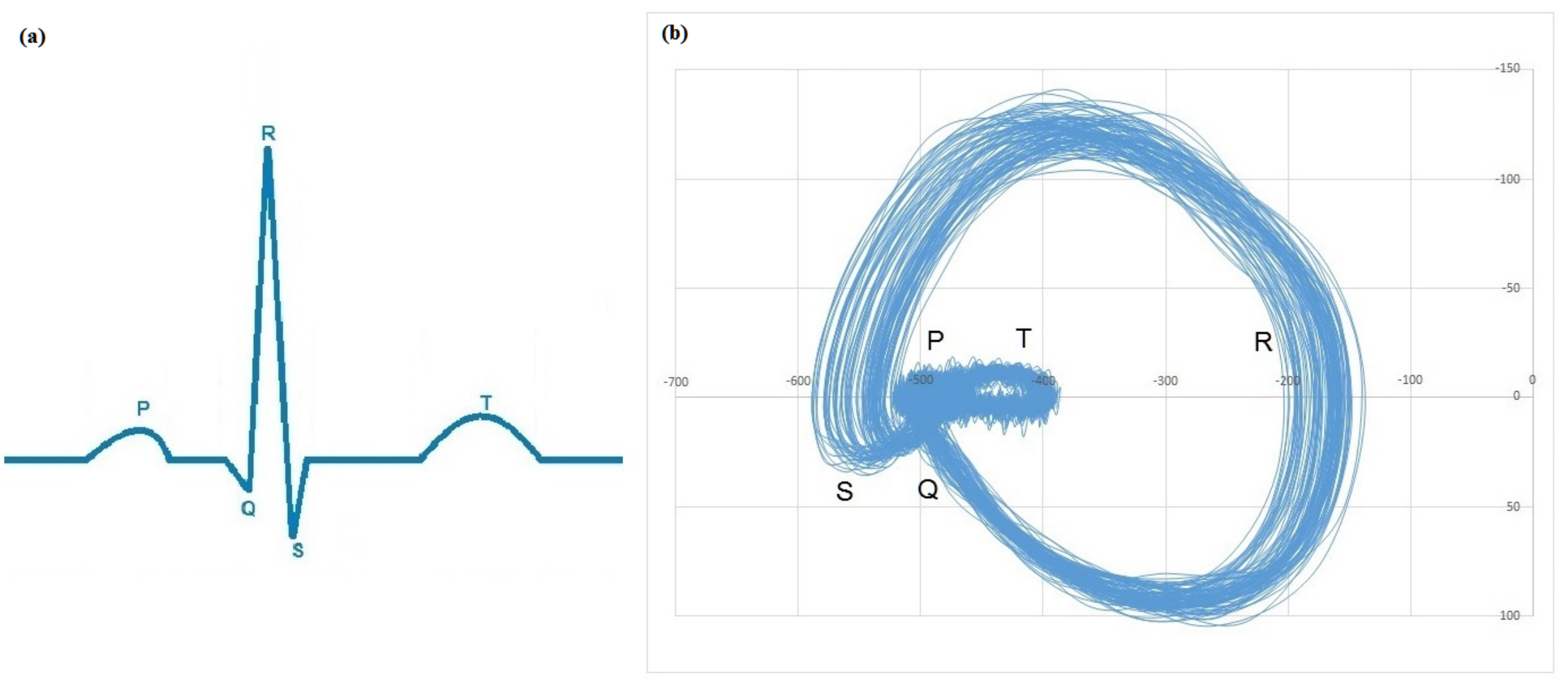}
        \caption{$PQRST$ in ECG and phase space. \textbf{(a)}	ECG \textbf{(b)} Phase space.}
	\label{Fig:Figure1}
\end{figure}

It is well known that human heart has an electric activity, which can be detected by measuring the potential difference from various points on the surface of the body. The measured electric potential versus time is called electrocardiogram (ECG), which possesses three separate parts. $P$-wave presents the excitation of the atria, $QRS$ complex shows the ventricles (His-Purkinje network) and $T$-wave is associated with the recovery of initial electrical state of ventricles (see Figure \ref{Fig:Figure1}(a)). Although ECG presents periodic behavior, some irregularities can be seen in details of the record. In fact, these irregularities belong to the intrinsic part of heart activity and/or to the random noise that can be found in such systems. These activities in ECG spectra are highly important to understand the cardiac dynamics. Cardiac oscillations are sometimes perturbed by unpredictable contributions which are part of the cardiac dynamics and therefore physiologically important. These finding predict that heart is not a perfect oscillator and/or cardiac muscles do not always vibrate harmonically.

It has been also well established that the transformation of a sequence of values in time to a geometrical object in space is a highly considered topic by \citet{Liebovitch1998}. This procedure replaces an analysis in time with an analysis in space. Here, the space is called phase space and the procedure of transforming the time series into the object in space is called an embedding. Afterwards, topological properties of the object are determined based on its fractal dimension. The fractal dimension characterizes the properties of the phase space set, not the original time series. The measured dimension in the phase space set determines whether a data set is generated by random or deterministic processes. A large value of the fractal dimension indicates random generation of the time series. This means that, the number of variables and equations are so large that there are no ways to predict the future values from the early parameters. In other words, the multiplicity of interacting factors precludes the possibility of understanding how the underlying mechanisms work. On the other hand, a low fractal dimension value indicates that the data is generated by deterministic mechanisms, based on a small number of independent variables, which helps to understand how the values in the past can be used to predict the values in the future. If the mechanism generating the data is deterministic, but the time series behaves like generated by random processes, then the system is considered as chaotic. A chaotic system is deterministic but not predictable in the long range. In a deterministic system which is not chaotic, the value of a variable at a given time can be used to generate the value of that variable at all times in future.

Box counting technique is generally used to determine the fractal dimension of the phase space, which is generated from the time series data. The values of fractal dimension give the number of independent values to construct the time series from which the phase space is generated. The box counting algorithm analyzes the phase space set generated from the points with coordinates  $x(t)$, $x(t + \Delta t)$, .  .  .  , $x(t + (n-1) \Delta t)$, where $x(t)$ are the time series values and $\Delta t$ is the lag.

At early times, electric activity of heart was modeled by using a set of nonlinear differential equations \citet{vanderPol,Katholi, West}. Afterwards, variations of cardiac spectra originated from deterministic dynamics called "chaotic dynamics" which is highly sensitive to heart’s initial conditions are analyzed by \citet{Babloyantz1985,Babloyantz1986}. Analyzing the power spectra of a normal human heart shows that His-Purkinje network possesses a fractal like structure in \citet{Goldberger1985} and the existence of chaotic behavior of heart can be found out from the phase space picture by evaluating a fractal dimension, $D$, where phase space trajectories are extracted from the time series graph of ECG (Figure \ref{Fig:Figure1}(b)) \citet{Babloyantz1988}. Lower values of $D$ indicate more coherent dynamics. If $D=1$, the oscillation is periodic and the phase space picture shows limiting cycle. However, $D$ becomes larger than one when a limiting cycle is perturbed by random noise \citet{Berge}. $D$ has non integer values greater than two when the system becomes chaotic or strange attractor \citet{Berge}. In this case, although trajectories in time do not converge towards a limiting cycle, they stay in a bounded region in the phase space. Instead of $D$, \citet{Babloyantz1988} evaluates the correlation dimension $D_2$ from a time series of finite length using the existing algorithms, \citet{Grassberger1,Grassberger2}. The produced correlation dimensions are obtained from a total of 36 ECG leads taken from 4 normal resting persons. Within the range of computational errors, they find values of $D_2$ ranging from $3.6$ to $5.2$. These values suggest that the normal cardiac oscillations follow a deterministic dynamics of chaotic nature.

It is also shown that a short-term heart rate variability analysis yields a prognostic value in risk stratification, independent of clinical and functional variables, \citet{LaRovere}. However, the detailed description and classification of dynamical changes using time and frequency measures are often not sufficient, especially in dynamical diseases as characterized by \citet{Mackey1977, Mackey1979}.

\citet{Wessel} try to answer the question: is the normal heart rate chaotic due to respiration? In their work, they give an example of the influence of respiration on heart beat dynamics, showing that observed fluctuations can mostly be explained by respiratory modulations of heart rate and blood pressure. Recently, the development of a fast and robust method which can be applied to multichannel physiologic signals was reported by \citet{Wilson}. This method elaborates either removing a selected interfering signal or separating signals that arise from temporally correlated and spatially distributed signals such as maternal or fetal ECG spectra. Convolutional neural networks (CNNs) method was also applied to patient specific ECG classification for real-time heart monitoring, \citet{Kiranyaz}.

Nowadays, it is well understood that cardiac diseases are one of the main reasons of mortality in modern, industrialized societies, and they cause high expenses in public health systems. Therefore, it is important to develop analytical methods to improve cardiac diagnostics. In this work, we investigate two different ECG systems taken from normal and abnormal human hearts, \cite{Goldberger2000}. Our aim is to introduce auxiliary phase space method in conjunction with ECG signals to diagnose heart diseases. We apply fractal analysis to the given data through trajectories produced in phase space, from where fractal dimension $D$ is obtained with the use of the box counting method, \cite{Liebovitch1989}.

\section{Methods}
\label{sec:headings}

\subsection{Data}
\begin{figure}[h!]
\begin{center}
\includegraphics[width=16.5cm]{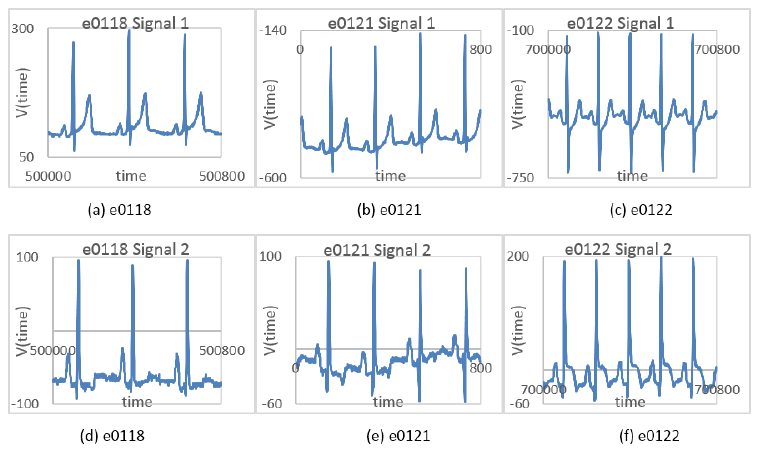}
\caption{First and Second ECG signal of normal heart taken at two different records of the same person.}
\label{Fig:Figure2}
\end{center}
\end{figure}

\begin{figure}[h!]
\begin{center}
\includegraphics[width=16.5cm]{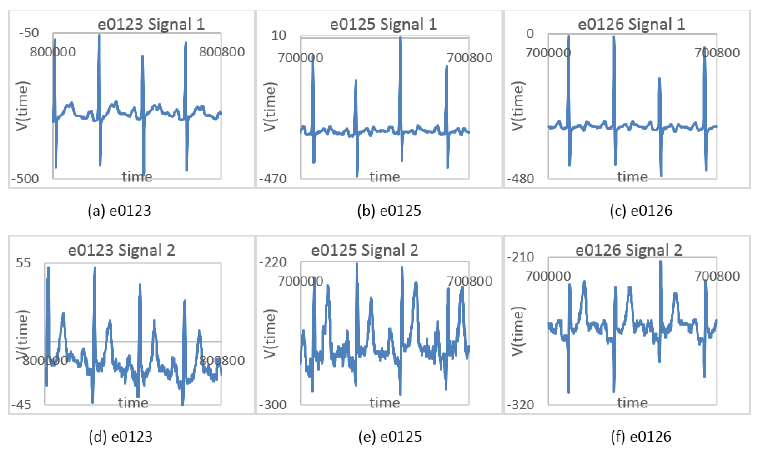}
\caption{First and Second ECG signal of abnormal heart taken at two different records of the same person.}
\label{Fig:Figure3}
\end{center}
\end{figure}

The data are taken from European $ST$-$T$ Database intended to be used for the evaluation of algorithms for analysis of $ST$ and $T$-wave changes, \cite{Goldberger2000}. We have selected three different ECG records of two persons. Person 1, considered as having a normal heart, is a man aged 51 with resting angina and normal coronary arteries. Person 2, considered as having an abnormal heart, is a man aged 58 with resting angina, anterior myocardial infarction, 1-vessel disease (LAD) and aortic valvular regurgitation. The ECG records e0118, e0121 and e0122 of person 1 (Figure \ref{Fig:Figure2}) and the ECG records e0123, e0125 and e0126 of person 2 (Figure \ref{Fig:Figure3}) are examined. Each record has two signals registered based on lead $V_4$ and modified lead III (MLIII), respectively. For each signal, 200,000 samples are used.

\subsection{Algorithms}
\subsubsection{Phase space}
\begin{figure}[h!]
\begin{center}
\includegraphics[width=13.1cm]{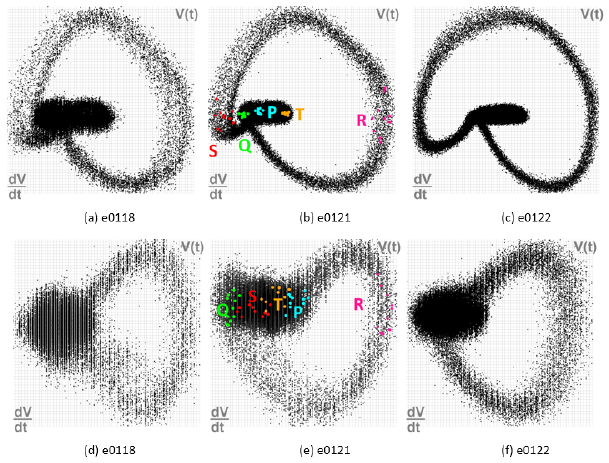}
\caption{Phase space of normal heart for the first and the second signal.}
\label{Fig:Figure4}
\end{center}
\end{figure}

\begin{figure}[h!]
\begin{center}
\includegraphics[width=13.1cm]{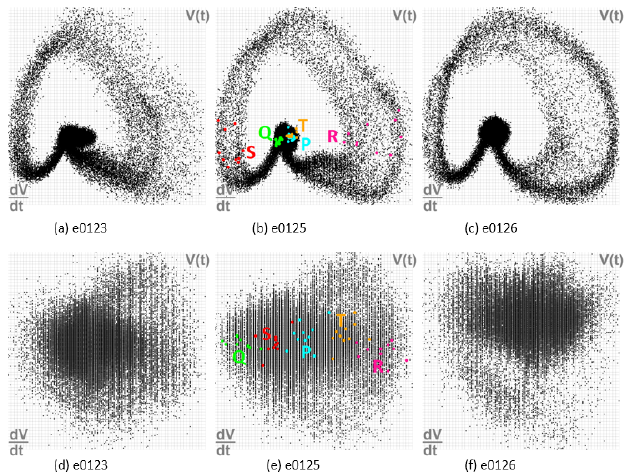}
\caption{Phase space of abnormal heart for the first and the second signal.}
\label{Fig:Figure5}
\end{center}
\end{figure}

We construct the phase space based on heart voltage values over time (i.e., $V(t)$) and their first derivative (i.e., $dV(t) / dt$). Figure \ref{Fig:Figure4}, and Figure \ref{Fig:Figure5} show the phase spaces, $V(t)$ versus $dV(t)/dt$.

To obtain the first derivative of the function $V(t)$, we use third-order forward difference Taylor Series derivative approximation of $f'(x)$ \citet{Burden, Khan, Ronco}. The simple approximation of the first derivative of a function $f$ at a point $x$ is defined as the limit of a difference quotient as follows:

\begin{equation}
f'(x)=\lim_{h \rightarrow 0} \frac {f(x+h)-f(x)}{h}
\end{equation}

If $h > 0$, meaning that $h$ is a finite positive number, then
\begin{equation}
f'(x)=\frac {f(x+h)-f(x)}{h}
\end{equation}
is called the first-order forward difference approximation of $f'(x)$. The approximation of $f'(x)$ can be obtained by combining Taylor series expansions. If the values ($x_{(i+1)}$,$f_{(i+1)}$), ($x_{(i+2)}$,$f_{(i+2)}$) and ($x_{(i+3)}$,$f_{(i+3)}$) are known, the first order derivative of $f_i(x)$ can be calculated as shown by $f'_i(x)$.
\begin{equation}
f_{(i+1)}=f_i+\frac{f'_i}{1!} h + \frac {f''_i}{2!} h^2 + \frac {f'''_i}{3!} h^3 +\ldots+\frac {f_i^n}{n!} h^n + R_n
\end{equation}
\begin{equation}
(x_{i+1},f_{i+1} ) \rightarrow  f_{i+1}=f_i + \frac {f'_i} {1!} h + \frac {f''_i}{2!} h^2
\rightarrow f_{i+1}=f_i+ h f'_i+ \frac {h^2}{2} f''_i
\end{equation}
\begin{equation}
(x_{i+2},f_{i+2} ) \rightarrow f_{i+2}=f_i+ \frac {f'_i}{1!} (2h)+ \frac {f''_i}{2!} (2h)^2
\rightarrow f_{i+2}=f_i + 2 h f'_i + 2 h^2 f''_i
\end{equation}
\begin{equation}
(x_{i+3},f_{i+3} ) \rightarrow  f_{i+3}=f_i+ \frac {f'_i}{1!} (3h) + \frac {f''_i}{2!} (3h)^2
\rightarrow f_{i+3}=f_i+ 3 h f'_i + \frac{9}{2} h^2 f''_i
\end{equation}

The second derivative $f''_i$ is canceled by adding Equations 4, 5 and 6 after multiplying them by $18$, $-9$, and $2$, respectively. Therefore, we obtain:
\begin{equation}
18 f_{i+1}- 9 f_{i+2}+ 2 f_{i+3}= 11 f_i + 6 h f'_i
\end{equation}

Finally, third-order forward difference Taylor series approximation of $f'(x)$ is obtained as follows:

\begin{equation}
f'_i=\frac{1}{6h}(-11 f_i+18 f_{i+1}-9 f_{i+2}+ 2 f_{i+3})
\end{equation}

\subsubsection{Box counting}
\begin{figure}[h!]
\begin{center}
\includegraphics[width=8.0cm]{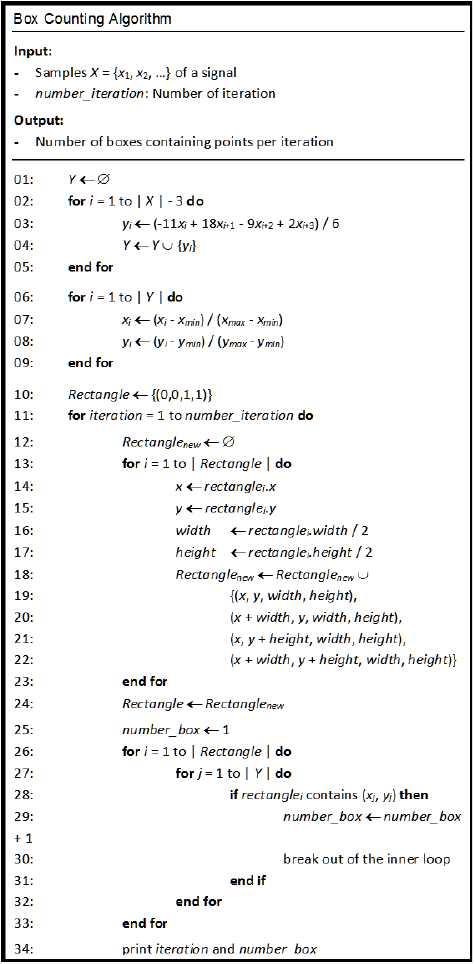}
\caption{Box Counting Algorithm.}
\label{Fig:Figure6}
\end{center}
\end{figure}

The pseudocode of the box counting algorithm is shown in Figure \ref{Fig:Figure6}. The samples of the signal constitute the $x$ coordinates of the time series. A $y$ coordinate is calculated based on four consecutive $x$ values as explained in the previous section (line 3). $x$ and $y$ values are then normalized (lines 6-9). A rectangle is indicated based on the ($x$,$y$) coordinates of its left top corner, width and height (e.g., line 10). The for loop in lines 11-35 gives the number of boxes containing points in the phase space. The algorithm starts with the smallest rectangle containing all points in the space (line 10). In each iteration, rectangles in the set Rectangle are divided into four rectangles (lines 12-24). In lines 25-33, we count the number of boxes containing at least one point in the space. The number of boxes per iteration are given at the end (line 34).

\section{Results and discussions}
Figure \ref{Fig:Figure4}, and \ref{Fig:Figure5} present the phase space pictures of ECG (time series) produced from Figure \ref{Fig:Figure2}, and Figure \ref{Fig:Figure3} for normal and abnormal hearts, respectively. Phase space pictures of normal hearts for the first signal in Figure \ref{Fig:Figure4} show deterministic behavior, meaning that this signal possesses a perfect oscillatory character.

When the phase space pictures in Figure \ref{Fig:Figure4} are compared with the phase space pictures of the first signal of the abnormal heart in Figure \ref{Fig:Figure5}, random behaviors are observed. Phase space pictures in Figure \ref{Fig:Figure5} imply that some perturbations start to develop on the top of the heart oscillations at the same time, which makes it difficult to understand the individual mechanisms that occur while producing the first signal of the abnormal heart. Figure 4 (d) to (f) present the second signal of the normal heart, which shows slight deviation from the oscillatory behavior, but still obeys the deterministic character. The broadening of this signal is most probably due to the resting angina which effects the heart oscillation. A similar broadening may be interpreted for the abnormal heart. However, as seen in Figure 5, all the phase space pictures have a strong random character, implying that the abnormal heart has serious heart failure due to anterior myocardial infarction, 1-vessel disease (LAD) and Aortic valvular regurgitation. In other words, in phase space terminology, numerous simultaneous factors are behind this random behavior of the abnormal heart. Here, in short, these phase space pictures can imply that the abnormal heart is unable to perform normal oscillatory action due to its cardiac deficiencies.

\begin{figure}[h!]
\begin{center}
\includegraphics[width=16.5cm]{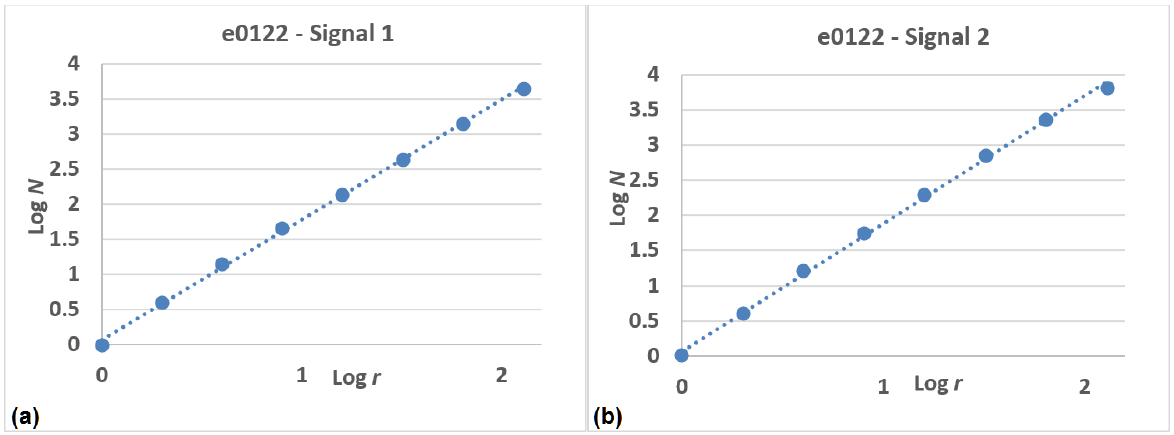}
\caption{$\log N$ versus $\log r$ plots and best fit for normal heart data.}
\label{Fig:Figure7}
\end{center}
\end{figure}

Figure \ref{Fig:Figure4}, and Figure \ref{Fig:Figure5} can also be interpreted with the help of the peak height, V(t) patterns in ECG in Figure \ref{Fig:Figure2},and Figure \ref{Fig:Figure3}. In phase space pictures, $R$ peaks always have larger values than $P$, $Q$, $S$ and $T$ peaks. On the other hand, $S$ peaks have the smallest values only in the first signal of both normal and abnormal hearts, while in the second signals $Q$ peaks have the smallest values compared to the other peaks. Moreover, the phase space picture in Figure \ref{Fig:Figure7}(b) provides well localized $R$, $P$, $T$, $S$ and $Q$ values, implying that peak heights of the ECG signals in Figure \ref{Fig:Figure2}(b) have almost the same values and are distinguished from each other by their location in the phase space picture. Spreading of $R$, $P$, $T$, $S$ and $Q$ values are more pronounced for the second signal’s phase space pictures in Figure \ref{Fig:Figure4}(e) due to a single defect, i.e. resting angina in normal heart. On the other hand, as seen in Figure \ref{Fig:Figure5}(a-b-c), especially the spreading of the R values reflects the randomness of peak heights of $R$ in the ECG pattern, implying a serious heart failure due to anterior myocardial infarction, 1-vessel disease (LAD) and Aortic valvular regurgitation of the abnormal heart. This randomness of the peak heights of $R$, $P$, $T$, $S$ and $Q$ tremendously increases for the second signal of the abnormal heart, as seen in Figure \ref{Fig:Figure5}(d-e-f), where $R$ and $P$, $T$, $Q$, $S$ values spread in all directions and become indistinguishable from each other in phase space, showing a strong random character. In that sense, one of the significant outcomes of our work is that the phase space analysis can be useful for the diagnosis of heart diseases in conjunction with ECG patterns, because the broadening of $R$, $P$, $T$, $S$ and $Q$ values can easily imply the irregularities in ECG patterns and provides information for peak height distribution.

\begin{figure}[h!]
\begin{center}
\includegraphics[width=16.0cm]{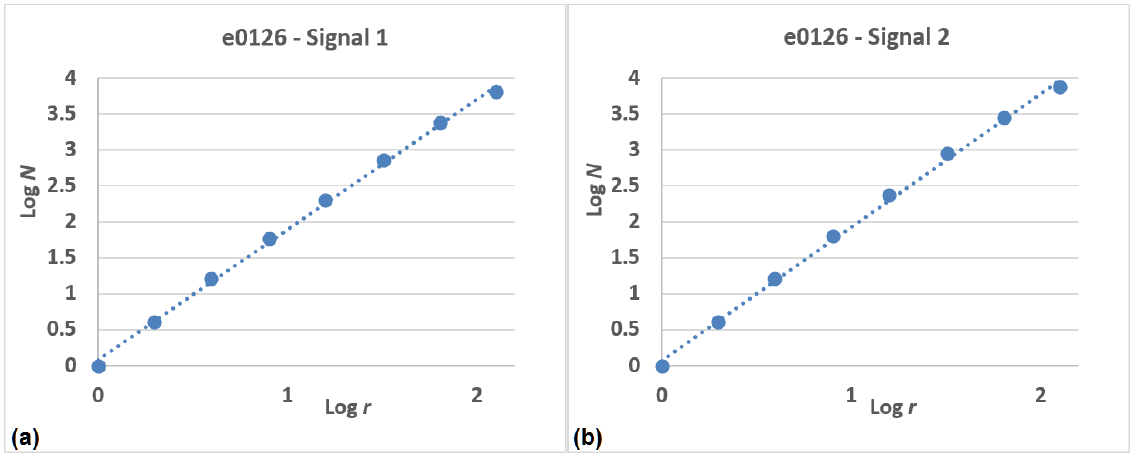}
\caption{$\log N$ versus $\log r$ plots and best fit for abnormal heart data.}
\label{Fig:Figure8}
\end{center}
\end{figure}

The data in Figure \ref{Fig:Figure4}, and Figure \ref{Fig:Figure5} are elaborated using box counting method in Figure \ref{Fig:Figure6}
, where the following equation (9) is used

\begin{equation}
D=\frac{\log N}{\log \displaystyle\frac {1}{r}}
\end{equation}

to produce fractal dimension, $D$. In this equation, $N$ is the minimum number of boxes needed to cover the set of points and $r$ is the box size. The plots of $\log N$ versus $\log r$ are given in Figure \ref{Fig:Figure7} and Figure \ref{Fig:Figure8}, from where $D$ values are calculated.

The results are shown in Table I together with the correlation coefficients, $R^2$ which are found out to be in a reasonable range.

Fractal dimensions, $D$, calculated from the phase space pictures in Figure \ref{Fig:Figure4}, and Figure \ref{Fig:Figure5} by using the box counting method are listed in Table I. $D$ values calculated from the normal heart data (i.e., 1.787, 1.749, 1.708 for the first signal and 1.804, 1.816, 1.821 for the second signal) are smaller than the abnormal heart dimensions (i.e., 1.816, 1.814, 1.816 for the first signal and 1.863, 1.861, 1.860 for the second signal), supporting the existing deterministic behavior of the normal heart compared to the random behavior of the abnormal heart. In other words, the former is a good oscillator, while the latter has some serious difficulties to make perfect harmonic oscillations.

\begin{table}[h!]
\caption{Fractal Dimensions Produced from Phase Space in Figure \ref{Fig:Figure4}, and Figure \ref{Fig:Figure5} and Fitting Procedures in Figure \ref{Fig:Figure7}, and Figure \ref{Fig:Figure8}.}
\begin{center}
\label{tab:1}       
\begin{tabular}{|l|l|ll|ll|}
\hline
 &  & Fr.Dim.$D$ & & Fitting $R^2$ & \\
\hline
 & Patient & Signal1 & Signal2 & Signal1 & Signal2 \\
\hline
Normal  & e0118 & 1.787 & 1.804 & 0.9981 & 0.9974\\
        & e0121 & 1.749 & 1.816 & 0.9992 & 0.9980\\
        & e0122 & 1.708 & 1.821 & 0.9992 & 0.9987\\
\hline
Abnormal& e0123 & 1.816 & 1.863 & 0.9978 & 0.9983\\
        & e0125 & 1.814 & 1.861 & 0.9982 & 0.9980\\
        & e0126 & 1.816 & 1.860 & 0.9982 & 0.9977\\
\hline
\end{tabular}
\end{center}
\end{table}

\section{Conclusion}
The most crucial observation in this study is the behavior of the second signals (MLIII) which gives more information than the first signals ($V_4$) for the action of normal and abnormal hearts. The second signals have larger fractal dimension $D$ values than the first signals, predicting more randomness yet more information about MLIII measurements. The lowest value of $D$ (i.e., 1.708) indicates a perfect oscillation of the heart and the highest value of $D$ (i.e., $1.863$) presents randomness of the heart. In fact, phase space picture presents the distribution of the peak heights in the ECG spectra, giving valuable information about heart activities in conjunction with ECG itself. In future work, we plan to apply, both fractal dimension and peak height distribution analysis in phase space for various abnormal human hearts to improve the novel diagnoses method for heart diseases.

\paragraph{} \textbf{Acknowledgment}

We would like to thank Dr. Eliya Buyukkaya from Wageningen University for visualizing the ECG data as well as for her fruitful discussions with us.

\paragraph{} \textbf{Conflict of interest}

The author declares no conflict of interest.

\paragraph{} \textbf{Author’s contribution}

The authors contributed equally to this work, and read and approved the final manuscript.

\paragraph{} \textbf{Availability of supporting data}

All data used in this paper are obtained from European ST-T Database. It is available at
\url{https://physionet.org/physiobank/database/edb/}.

\paragraph{} \textbf{Ethical approval and consent to participate}

Kadir Has University sees no Ethical conflict about this study and approves this manuscript to be submitted to scientific journals. Besides the need for University consent was waived.

\bibliographystyle{unsrtnat}
\bibliography{references}  






\end{document}